\documentclass[letterpaper,twocolumn,showpacs,prb,aps]{revtex4-1}

\usepackage{amsmath}  
\usepackage{amsfonts} 
\usepackage{graphicx} 
\usepackage{amssymb}

\begin{document}

\title{Relaxation of charge in monolayer graphene:  fast non-linear diffusion vs Coulomb effects}

\author{Eugene B. Kolomeisky$^{1}$ and Joseph P. Straley$^{2}$}

\affiliation
{$^{1}$Department of Physics, University of Virginia, P. O. Box 400714,
Charlottesville, Virginia 22904-4714, USA\\
$^{2}$Department of Physics and Astronomy, University of Kentucky,
Lexington, Kentucky 40506-0055, USA}

\date{\today}

\begin{abstract}
Pristine monolayer graphene exhibits very poor screening because the density of states vanishes at the Dirac point.  As a result, charge relaxation is controlled by the effects of zero-point motion (rather than by the Coulomb interaction) over a wide range of parameters.  Combined with the fact that graphene possesses finite intrinsic conductivity,  this leads to a regime of relaxation described by a non-linear diffusion equation with a diffusion coefficient that diverges at zero charge density.  Some consequences of this fast diffusion are self-similar superdiffusive regimes of relaxation, the development of a charge depleted region at the interface between electron- and hole-rich regions, and finite extinction times for periodic charge profiles.   
       
\end{abstract}

\pacs{72.80.Vp, 73.22.Pr, 73.50.Gr}

\maketitle

\section{Introduction}

Understanding how excess charge carriers approach equilibrium is important to both condensed matter and device physics.  In an isotropic three-dimensional medium of electrical conductivity $\sigma$ and  dielectric constant $\kappa$ the driving force of relaxation is the Coulomb interaction; relaxation is exponentially fast and characterized by the Maxwell relaxation rate $4\pi\sigma/\kappa$.  In earlier pioneering studies Dyakonov and Furman, and Govorov and Chaplik (DFGC) \cite{DFGC}  have recognized that the Maxwell relaxation law does not apply to low-dimensional systems. In the two-dimensional case relevant to semiconductor heterostructures \cite{AFS}, the two-dimensionally confined charges interact according to the three-dimensional Coulomb law;  then the counterpart of the Maxwell relaxation rate ($2\pi\sigma/\kappa$) is dimensionally a velocity, suggesting that the charged region expands with constant velocity proportional to $\sigma/\kappa$ \cite{DFGC}.  Non-Maxwellian relaxation in two-dimensional systems has been experimentally detected \cite{exp}.   

Here we will discuss the relaxation of charge in monolayer graphene \cite{graphene_review}.  This problem was posed by Efros \cite{Efros}, who conjectured that Coulomb forces continue to dominate the physics, so that the DFGC theory \cite{DFGC} would apply.  However, we will find that in the low doping limit the situation is somewhat delicate.  Graphene is distinguished from other two-dimensional electron systems in that reduction in doping causes a decrease in the density of states (DOS).  This has its origin in the pseudo-relativistic dispersion law of graphene's elementary excitations \cite{graphene_review}, and translates \cite{AFS} into a decreasing screening response as the Dirac point is approached.  This can be quantified as an increase in the Debye screening length with decrease in doping.  Screening is an important consideration here because the Coulomb interaction  dominates relaxation only on the scale exceeding the screening length. When the screening length is large compared to other characteristic scales of the problem, quantum-mechanical effects can take over.  This situation is most pronounced at the Dirac point, where the DOS vanishes and the screening length is infinite.  Here in a relevant range of parameters we identify a novel regime of relaxation described by a non-linear diffusion equation of the type first obtained in theories of thermal waves and filtration \cite{shock} and later encountered in studies of dynamics of crystal interfaces below the roughening phase transition \cite{Spohn}, soliton dynamics in one-dimensional conductors \cite{soliton}, diffusion in a plasma \cite{plasma}, spreading of liquid drops in the presence of Van der Waals interactions \cite{Joanny}, and self-organized criticality \cite{SOC}.  Relaxation in graphene has more in common with the last three examples as they all are characterized by having a divergent diffusion coefficient in the limit that the density of the diffusing quantity vanishes.  In graphene the effect of the "fast" non-linear diffusion manifests itself in the self-similar superdiffusive regimes of relaxation,  the development of charge depleted regions at the interface between electron-rich and hole-rich regions, and the finite extinction times of periodic charge profiles.   Recently demonstrated high-resolution non-invasive imaging of charge currents in graphene structures \cite{imaging} can be employed to observe these effects.

Below we focus on the interesting case of monolayer graphene at the Dirac point.   Our continuum theory is limited to zero temperature and applicable in the long-wavelength, low-frequency limit.  We additionally neglect the effects of retardation as all the relevant velocity scales that we will encounter are significantly smaller than the speed of light.  Whenever it cannot cause confusion, we also suppress the time dependence of the dynamical variables.  

\section{Statement of the problem}

The total potential $\varphi(\textbf{r})$ felt by a carrier of charge $e$ at a position $\textbf{r}$ is due to the potential of the other carriers of number density $n(\textbf{r})$
 \begin{equation}
\label{scpotential}
\varphi(\textbf{r})=\frac{e}{\kappa}\int \frac{n(\textbf{r}')}{|\textbf{r}-\textbf{r}'|}d^{2}r'
\end{equation}
where $\kappa$ is due to graphene's own electrons and surrounding environment.  In thermodynamic equilibrium the electrochemical potential
\begin{equation}
\label{equilibrium}
\mu=\zeta(n)+e\varphi(\textbf{r})=\hbar v_{F}\sqrt{\pi n}+e\varphi(\textbf{r})
\end{equation}
is zero across the system.  Here $\zeta(n)=\hbar v_{F}\sqrt{\pi n}$ is the chemical potential for charge carriers in the absence of a perturbing potential $\varphi(\textbf{r})$, $v_{F}$ is the Fermi velocity, and the density $n$ is assumed to be small enough that the low-energy Dirac dispersion law, $\varepsilon =\hbar v_{F}k$, holds \cite{graphene_review}.   The equilibrium state is carrier-free, $n=0$, and $\varphi=0$. 

Out of equilibrium the driving force for evolution is the negative of the gradient of the electrochemical potential:
\begin{equation}
\label{driving_force}
\textbf{F}=-\nabla \mu=-\frac{\partial \zeta}{\partial n}\nabla n-\frac{e^{2}}{\kappa}\nabla \int\frac{n(\textbf{r}')d^{2}r'}{|\textbf{r}-\textbf{r}'|}
\end{equation}
where the first term, inversely proportional to the DOS $\partial n/\partial\zeta$, is the consequence of Fermi statistics (increasing density causes an increase in the total kinetic energy of the charge carriers), while the second term is the Coulomb force.  The DOS determines the Debye screening length of the two-dimensional electron gas \cite{AFS}
\begin{equation}
\label{screening_length}
q_{s}^{-1}(n)=\frac{\kappa}{2\pi e^{2}}\frac{\partial \zeta}{\partial n}=\frac{1}{4\alpha\sqrt{\pi n}},~~\alpha=\frac{e^{2}}{\kappa \hbar v_{F}}\approx \frac{2.5}{\kappa}
\end{equation} 
which diverges as $n\rightarrow 0$.  Here $\alpha$ is the fine structure constant for graphene \cite{graphene_review};  throughout this work we assume that $\alpha\ll1$ or equivalently $\kappa\gg1$.  Suspended graphene, characterized by $\alpha \approx 1/2$ \cite{suspended}, is at the verge of applicability of our theory.   

According to Ohm's law the force (\ref{driving_force}) induces an electric current density 
\begin{equation}
\label{Ohm}
\textbf{j}=\frac{\sigma_{0}}{e}\textbf{F}=\frac{4e}{\pi\hbar}\textbf{F}
\end{equation}
where the conductivity of graphene $\sigma_{0}$ is approximated by its intrinsic value $\sigma(n\rightarrow 0)= 4e^{2}/\pi \hbar$ \cite{Kats}.  Conservation of charge within a region implies a continuity equation
\begin{equation}
\label{continuity}
e\frac{\partial n}{\partial t}+\nabla \textbf{j}=0
\end{equation}

\section{Relaxation dominated by Coulomb effects}

The quantum-mechanical effects accumulated in the first term of the expression for the driving force of relaxation (\ref{driving_force}) can be neglected when the condition  
\begin{equation}
\label{dense}
\alpha L\sqrt{n}\gg 1 
\end{equation}             
holds.  This requires that the characteristic length scale of the problem $L$ be significantly larger than the local Debye screening length $q_{s}^{-1}(n)$ (\ref{screening_length}).  Then the theory is linear and can be solved by a Fourier transform;  it is equivalent to the DFGC theory \cite{DFGC}, whose hallmark is the linear evolution equation     
\begin{equation}
\label{DFGC_real_space}
\frac{\partial^{2}n}{\partial t^{2}}+ v_{0}^{2}\triangle n=0,~~v_{0}=\frac{2\pi \sigma_{0}}{\kappa}=\frac{8e^{2}}{\kappa \hbar}=8\alpha v_{F}
\end{equation}
where the velocity scale  $v_{0}$ owes its existence to the intrinsic conductivity of graphene \cite{Kats}. We now quote two solutions \cite{DFGC} of Eq.(\ref{DFGC_real_space}) supplemented with their ranges of applicability according to Eq.(\ref{dense}).  For an initially localized one-dimensional distribution of number density $N=\int_{-\infty}^{\infty}n(x,0)dx$ (a stripe of excess charge), the large-time asymptotic solution to Eq.(\ref{DFGC_real_space}) has the form
\begin{equation}
\label{DFGC_1d_solution}
n(x,t)=\frac{N}{\pi}\frac{v_{0}t}{x^{2}+(v_{0}t)^{2}}, ~~~t\gg t_{x}\simeq \frac{1}{\alpha^{3}Nv_{F}}
\end{equation}
while for an initially localized two-dimensional distribution of strength $Z=\int n(\textbf{r},0)d^{2}r$, one similarly finds 
\begin{equation}
\label{DFGC_2d_solution }
n(\textbf{r},t)=\frac{Z}{2\pi}\frac{v_{0}t}{[r^{2}+(v_{0}t)^{2}]^{3/2}}, ~~~Z\gg Z_{x}\simeq\frac{1}{\alpha^{2}}
\end{equation}
These results show that most of the charge in the expanding clouds is localized within a distance $L\approx v_{0}t$ from the origin; the clouds spread with the constant velocity $v_{0}$.  Eq.(9) applies at times exceeding the cross-over time scale $t_{x}$ which can be very large for $N$ small.  In this case the intermediate asymptotic behavior for $t\ll t_{x}$ is dominated by quantum-mechanical effects.  Similarly, Eq.(\ref{DFGC_2d_solution }) fails to describe relaxation of clouds of net dimensionless charge $Z$ smaller than the crossover charge $Z_{x}$.   For example, for $\alpha=1/10$ one finds $Z_{x}\simeq100$.  Excess charges smaller than a $100$ are certainly easier to realize in practice than charges in excess of it.   Existence of a wide range of parameters where quantum-mechanical effects dominate relaxation of charge in graphene is a direct consequence of the vanishing DOS, $\partial n/\partial \zeta \propto \sqrt{n}\rightarrow 0$ as $n\rightarrow 0$.  

\section{Relaxation dominated by quantum effects}

Having established the range of applicability of the DFGC theory \cite{DFGC} to relaxation in graphene, we now turn to the regime when physics is dominated by quantum-mechanical effects, and assume that the condition opposite to Eq.(\ref{dense}) holds.

Neglecting the Coulomb term in Eq.(\ref{driving_force}), the force exerted on a carrier causes an electric current (\ref{Ohm})
\begin{equation}
\label{current_Dirac_point}
\textbf{j}=-eD(n)\nabla n, 
\end{equation}
where we have introduced the diffusion coefficient
\begin{equation}
\label{diffusion_constant}
D(n)=\frac{\sigma_{0}}{e^{2}}\frac{\partial \zeta}{\partial n}=\frac{2v_{F}}{\sqrt{\pi n}}
\end{equation}
The first representation is the Einstein relation, which makes it clear that the $n^{-1/2}$ divergence of $D(n)$ is the consequence of the finite intrinsic conductivity of graphene \cite{Kats} and of the vanishing DOS at the Dirac point \cite{graphene_review}.  Curiously, Planck's constant which is present in both the conductivity ($\sigma_{0}=4e^{2}/\pi \hbar$ \cite{Kats}) and in the inverse DOS, $\partial \zeta/\partial n\propto \hbar$, cancels out.  Thus the outcomes (\ref{current_Dirac_point}) and (\ref{diffusion_constant}) are independent of $\hbar$. 

\subsection{Steady state in the presence of fixed current}
 
The simplest non-equilibrium effect to consider is the steady state in the presence of a fixed current $\textbf{j}$ flowing through the system. This can be due to the hole flow along $\textbf{j}$, the electron flow opposite to $\textbf{j}$ or both flows present at once and meeting at a line annihilation front that is perpendicular to $\textbf{j}$.  Focusing on the last case, choosing the direction of $\textbf{j}$ to coincide with the negative $x$ axis the charge density distribution $\rho(x)=en(x)$ can then be found by integrating Eq.(\ref{current_Dirac_point}):
\begin{equation}
\label{st_charge_profile}
\rho(x>0)=\frac{\pi j^{2}}{16ev_{F}^{2}}x^{2},~~~\rho(x<0)=-\rho(x>0)
\end{equation} 
where the holes ($e>0$) present in the $x>0$ region flow toward $x=0$ where they annihilate with the electrons present at $x<0$ and flowing toward $x=0$.  Since graphene is a gapless material, there is no energy released during  the event of the electon-hole annihilation, and thus there is no additional force associated with the process.  The $ x^{2}$ behavior of the density is a necessary outcome of having a finite current density at the annihilation line $x=0$, made possible by the interplay between the diffusion coefficient (\ref{diffusion_constant}) (which diverges as $1/x$) and the gradient of the density in Eq.(\ref{current_Dirac_point}) (which vanishes as $x$).  The rapid $x^{2}$ drop in density near the annihilation front is an example of a charge depletion region or a sink brought about by fast singular diffusion.  The result (\ref{st_charge_profile}) is valid within a range $|x|\lesssim L$ satisfying the condition $\alpha L^{2}j/v_{F}e\ll 1$.  

\subsection{Fast diffusion equation}
        
Substituting the expression for the current density (\ref{current_Dirac_point}) into the continuity equation (\ref{continuity}) and employing Eq.(\ref{diffusion_constant}) we arrive at our central result
\begin{equation}
\label{nl_diffusion_eq}
\frac{\partial n}{\partial t}=\nabla[D(n)\nabla n]=\frac{2v_{F}}{\sqrt{\pi}}\nabla(n^{-1/2}\nabla n).
\end{equation}     
This is a non-linear diffusion equation with a diffusion coefficient that diverges as $n\rightarrow 0$.  Not only does it not contain Planck's constant, but it also does not depend on the electron charge or the dielectric constant $\kappa$,  because the equation describes the regime where the Coulomb interaction is negligible.  The only material parameter entering Eq.(\ref{nl_diffusion_eq}) is the Fermi velocity $v_{F}$.  The $n^{-1/2}$ singularity of the diffusion coefficient is the same as that found in the problem of diffusion of plasma particles in the presence of magnetic field of a toroidal multipole \cite{plasma}.  

The remarkable property of the non-linear diffusion equation with a power-law dependence $D(n)$ is the existence (for a narrow class of problems) of exact self-similar solutions capturing the asymptotic large-time behavior of more general problems \cite{shock}.  The cases of "slow", $D(n\rightarrow 0)\rightarrow 0$, and "fast", $D(n\rightarrow 0)\rightarrow \infty$, diffusion are qualitatively different, and graphene belongs to the latter category.   Below we analyze a few particular cases of late stage relaxation in graphene which illustrate the salient features of fast diffusion.  These can be understood in simple terms via a combination of scaling arguments and conservation laws \cite{shock}, and are likely to be realized in practice.  

\subsubsection{One-dimensional localized charge distribution}

When the initial and/or boundary conditions depend only on one coordinate,  Eq.(\ref{nl_diffusion_eq}) becomes 
\begin{equation}
\label{1d_diffusion_eq}
\frac{\partial n}{\partial t}=\frac{2v_{F}}{\sqrt{\pi}}\frac{\partial}{\partial x}\left (n^{-1/2}\frac{\partial n}{\partial x}\right )
\end{equation}
An initially localized distribution (such as $\int_{-\infty}^{\infty}n(x,0)dx=N$) remains localized through the evolution.  Then dimensional considerations determine $L$, the width of the region where most of the charge is localized at a time $t$,  to have the form $L^{2}\simeq D(n)t \simeq v_{F}t/\sqrt{n}$.  Combining this with the estimate $nL\simeq N$ following from conservation of linear charge density ($N=\int_{-\infty}^{\infty}n(x,t)dx$), we find     
\begin{equation}
\label{1d_cloud _size}
L(t)\approx 2 N^{-1/3}(3v_{F}t)^{2/3}, ~~~t\ll t_{x}
\end{equation}
where the displayed numerical factors hereafter correspond to the asymptotically exact propagation speed of the charge wave $ dL/dt$ inferred from the exact solutions given below.  We see that the charged region expands with time faster than $t^{1/2}$ (linear diffusion) because the diffusion coefficient (\ref{diffusion_constant}) grows in the direction of spreading.  It is straightforward to verify that at the cross-over time scale $t\simeq t_{x}$, Eq.(\ref{DFGC_1d_solution}), the size of the cloud (\ref{1d_cloud _size}) has the same order of magnitude as that supplied by the DFGC theory, $v_{0}t_{x}$.  In the bulk of the distribution ($|x|\ll L(t)$), the density decreases with time as $n(0,t)\simeq N/L(t)\simeq N^{4/3}(v_{F}t)^{-2/3}$; well outside it ($|x|\gg L(t)$), the density can be directly estimated from Eq.(\ref{1d_diffusion_eq}) as $n/t\simeq v_{F}\sqrt{n}/x^{2}$ or $n(x,t)\simeq (v_{F}t)^{2}/x^{4}$.  The combination of these observations implies that the solution to Eq.(\ref{1d_diffusion_eq}) has the self-similar form
\begin{equation}
\label{Barenblatt1}
n(x,t)=\frac{N}{L(t)}f\left (\frac{x}{L(t)}\right )
\end{equation}
where the scaling function $f(\xi)$ satisfies the condition of conservation of charge density ($\int_{-\infty}^{\infty}f(\xi)d\xi=1$) and behaves as $f'(\xi\rightarrow 0)=0$ and $f(\xi\rightarrow \infty)\simeq 1/\xi^{4}$.  The exact result $f(\xi)=2/\pi(\xi^{2}+1)^{2}$ due to Zel'dovich, Kompaneets, and Barenblatt \cite{shock} exhibits these features.

\subsubsection{Two-dimensional localized charge distribution}  

For an initially localized two-dimensional distribution ($Z=\int n(\textbf{r},0)d^{2}r$), similar arguments hold except that we seek a solution to the radially-symmetric version of Eq.(\ref{nl_diffusion_eq})
\begin{equation}
\label{2d_diffusion_equation}
\frac{\partial n}{\partial t}=\frac{2v_{F}}{\sqrt{\pi}}\frac{1}{r}\frac{\partial}{\partial r}\left (rn^{-1/2}\frac{\partial n}{\partial r}\right ),
\end{equation}
and conservation of charge ($Z=\int n(\textbf{r},t)d^{2}r$) now implies the estimate $nL^{2}\simeq Z$.  As a result, for the size of the charge cloud at large time we find
\begin{equation}
\label{2d_cloud_size}
L(t)\approx \frac{8v_{F}t}{\sqrt{Z}}, ~~~Z\ll Z_{x}
\end{equation}    
We see that the charged region spreads out with the constant velocity $8v_{F}/\sqrt{Z}$ significantly exceeding that of the DFGC theory, $v_{0}=8\alpha v_{F}$, Eq.(\ref{DFGC_real_space});  both velocities have the same order of magnitude at the cross-over charge $Z_{x}$.  Moreover the expansion is faster than in one dimension (\ref{1d_cloud _size}) because the diffusion coefficient grows in all directions away from the cloud center.  Ballistic spreading with a velocity proportional to $v_{F}$ can be anticipated from the outset because $Z$ is dimensionless while $v_{F}$ is the only parameter of the problem having dimensionality (of velocity).  For a localized two-dimensional distribution the solution to Eq.(\ref{2d_diffusion_equation}) has the self-similar form  
\begin{equation}
\label{Barenblatt2}
n(\textbf{r},t)=\frac{Z}{L^{2}(t)}g\left (\frac{r}{L(t)}\right )
\end{equation}   
where the scaling function $g(\xi)$ satisfies the condition of conservation of charge ($2\pi \int_{0}^{\infty}g(\xi)\xi d\xi=1$) and behaves as $g'(\xi\rightarrow 0)=0$ and $g(\xi\rightarrow \infty)\simeq 1/\xi^{4}$.  The exact solution $g(\xi)=1/\pi(\xi^{2}+1)^{2}$, also due to Zel'dovich, Kompaneets, and Barenblatt \cite{shock}, exhibits these features.    

\subsubsection{One-dimensional localized dipole charge distribution}

Relaxation of one-dimensional charge profiles can be further accelerated compared to the $t^{2/3}$ result (\ref{1d_cloud _size}) by introducing a sink into the problem.  Indeed, let us consider a charge distribution composed of electrons ($x<0$) and holes ($x>0$) with localized number density, so that the total charge is zero while the first (dipole) moment $\int_{0}^{\infty}xn(x,0)dx=P$ is finite.  Due to annihilation events at $x=0$, the total number of electrons or holes is no longer conserved; however, the dipole moment $P= \int_{0}^{\infty}xn(x,t)dx$ is conserved by the   
equation of motion (\ref{1d_diffusion_eq}) \cite{shock}.  This conservation law implies the estimate $P\simeq nL^{2}$.  When combined with the relationship $L^{2}\simeq v_{F}t/\sqrt{n}$, we find
\begin{equation}
\label{ dipole_cloud_size}
L(t)\approx \frac{4\sqrt{2}}{\sqrt[4]{3}}\frac{v_{F}t}{\sqrt{P}}, ~~~P\ll Z_{x}
\end{equation}  
We see that similar to the case of a localized two-dimensional distribution (Eq.(\ref{2d_cloud_size})), the region occupied by carriers of each sign expands ballistically with a velocity of the order $v_{F}/\sqrt{P}$.  The total number of carriers (per unit length) of a given kind decreases with time as $nL\simeq P^{3}/v_{F}t$.  The analytic solution in this case has a self-similar form resembling Eq.(\ref{Barenblatt2}),
\begin{equation}
\label{BZ}
n(x,t)=\frac{P}{L^{2}(t)}h\left (\frac{x}{L(t)}\right ),
\end{equation}
where the scaling function $h(\xi\geqslant0)$ satisfies the condition of conservation of dipole moment($\int_{0}^{\infty}\xi h(\xi)d\xi=1$) and behaves as $h(\xi\rightarrow \infty)\simeq 1/\xi^{4}$ and $h(\xi\rightarrow 0)\simeq \xi^{2}$.  The latter property follows from the observation that the sink at $x=0$ only differs from its steady-state counterpart (\ref{st_charge_profile}) by time-dependent current $j(0, t)$.  The exact solution in this case, $h(\xi)=9\sqrt{3}\xi^{2}/2\pi(1+\xi^{3})^{2}$  (a relative of the "dipole" solution due to Barenblatt and Zel'dovich for the slow diffusion case \cite{shock}) exhibits these properties.

\subsubsection{One-dimensional periodic charge distribution}

One-dimensional relaxation can be drastically accelerated in the presence of a series of sinks.  A relevant example is a periodic charge profile with both electrons and holes present in equal amounts.  The spatial periodicity of the distribution is  preserved by time evolution which means that only a half-period occupied by carriers of one sign needs to be considered.  We further assume that the evolving density profile is symmetric about its maximum at $x=L$, where the current density (\ref{current_Dirac_point}) vanishes.  It is then sufficient to solve Eq.(\ref{1d_diffusion_eq}) in the $0\leqslant x \leqslant L$ quarter-period range subject to the boundary conditions $n(0,t)=0$, $\partial n(x,t)/\partial x|_{x=L}=0$, and then employ symmetry to extend the results beyond it.  If the initial density profile has amplitude $n_{0}$, position is measured in units of $L$ and time in units of the characteristic diffusion time $\sqrt{\pi n_{0}}L^{2}/2v_{F}$, we seek a solution to the problem in the form $n(x,t)=n_{0}\nu (x/L,2v_{F}t/L^{2}\sqrt{\pi n_{0}})$ where the dimensionless density $\nu(\xi,\tau)$ evolves according to
\begin{equation}
\label{periodic_problem}
\dot \nu=2(\nu^{1/2})'',~~~\nu(0,\tau)=0~~~\nu'({1,\tau})=0
\end{equation}     
Assuming a separable form for the solution ($\nu(\xi,\tau)=a(\tau)s(\xi)$),  the amplitude $a(\tau)$ and shape $s(\xi)$ functions satisfy the equations
\begin{equation}
\label{separability}
\frac{\dot a}{a^{1/2}}=2\frac{(s^{1/2})''}{s}=-\lambda
\end{equation}               
where $\lambda\geqslant0$ is a separation constant;  without loss of generality we can set $s(1)=1$ \cite{plasma}.  The solution for the amplitude is then $a(\tau \leqslant \tau_{ext})=\lambda^{2}(\tau_{ext}-\tau)^{2}/4$, and $a(\tau>\tau_{ext})=0$, thus implying that the charge profile goes extinct in a finite time $\tau_{ext}\simeq 1$.  This corresponds to the diffusion time scale $\sqrt{n_{0}}L^{2}/v_{F}$ in physical units, and represents another manifestation of fast diffusion.  The equation for $u = s^{1/2}$ can be integrated; it is similar to the Newtonian motion of a particle of unit mass that has position $u$ at time $\xi$ in a potential $V = \lambda u^{3}/6$:
\begin{equation}
\label{quadrature}
\int_{0}^{s^{1/2}(\xi)}\frac{du}{(1-u^{3})^{1/2}}=\sqrt{\frac{\lambda }{3}}\xi, ~\lambda=B^{2} (1/2,1/3)/3,  
\end{equation}
where $B(x,y)$ is Euler's beta function, and the value of $\lambda\approx5.898$ is fixed by evaluation of the integral at $\xi=1$.  Collecting these observations, the solution to the problem (\ref{periodic_problem}) is
\begin{equation}
\label{solution}
\nu(\xi,\tau)\approx 8.698(\tau_{ext}-\tau)^{2}s(\xi) 
\end{equation}  
for $\tau \leqslant \tau_{ext}$, and $\nu=0$ for $\tau>\tau_{ext}$.  It has been proven that an arbitrary initial profile $\nu(\xi,0)$ on an interval satisfying Dirichlet boundary conditions evolves into the separable form (\ref{solution}) and rigorous bounds on $\tau_{ext}$ in terms of $\nu(\xi,0)$ can be given \cite{plasma}. Its implication for relaxation of periodic charge profiles in graphene is that after an initial transient they all "settle" into the fixed shape $s(\xi)$, Eq.(\ref{quadrature}), with amplitude falling off as $(\tau_{ext}-\tau)^{2}$;  relaxation ends in a finite time.   It is straightforward to infer from Eq.(\ref{quadrature}) that near the $\xi=0$ sink the profile exhibits the already familiar $\xi^{2}$ density drop.  Near its maximum the shape function is parabolic, i. e. $1-s(\xi\rightarrow 1)\propto (1-\xi)^{2}$.  

\section{Interplay of Coulomb and quantum effects at finite doping}

Quantum-mechanical effects have their strongest influence on relaxation at zero doping, but they continue to play a role at finite doping $n_{0}$.  When the deviation $\delta n(x,t)=n(x,t)-n_{0}$ is small, a linear theory captures the physics of relaxation.  It is then straightforward to show that small-wavevector $q\ll q_{s}(n_{0})$ modes of $\delta n$ evolve according to the DFGC equation (\ref{DFGC_real_space}) while large-wavevector $q\gg q_{s}(n_{0})$ perturbations relax according to the linear diffusion equation with the diffusion constant $D(n_{0})$, Eq.(\ref{diffusion_constant}).  The divergence of the screening length $q_{s}^{-1}(n_{0})$ (\ref{screening_length}) as $n_{0}\rightarrow 0$ is a  clear indicator that quantum-mechanical effects control relaxation of a wide range of modes. 

\section{Acknowledgement}

We are grateful to Eva Andrei for a valuable discussion of the topics covered in this work.

\end{document}